\documentclass[11pt]{article}
\usepackage{aaspp4}

\makeatletter

\let\@internalcite\cite
\def\cite{\def\citename##1{##1}\@internalcite}
\def\shortcite{\def\citename##1{}\@internalcite}

\def\reserved@a{LaTeX2e}
\ifx\reserved@a\fmtname
  \usepackage{epsfig}
  \newcommand{\myepsf}[3]%
{\begin{center}\epsfig{file=#2,height=#1,angle=#3}\end{center}}
  \newcommand{\myepsfbb}[7]%
{\begin{center}\epsfig{file=#2,height=#1,angle=#3,%
bbllx=#4,bblly=#5,bburx=#6,bbury=#7}\end{center}}
\else
  \newcommand{\myepsf}[3]%
{\begin{center}\epsfysize=#1 \leavevmode\epsfbox{#2}\end{center}}
  \newcommand{\myepsfbb}[7]%
{\begin{center}\epsfysize=#1 \leavevmode%
\epsfbox[#4 #5 #6 #7]{#2}\end{center}}
\fi

\makeatother

\newcommand{\etotal}{e_{\rm tot}}
\newcommand{\machrecip}{M_A^{-1}}
\newcommand{\machrecipcrit}{M_{A,i}^{-1}}
\newcommand{\machrecipfund}{M_{A,F}^{-1}}
\newcommand{\machrecipover}{M_{A,O}^{-1}}
\newcommand{\machreciptrans}{{{\cal M}_A^{-1}}}
\newcommand{\rhoIN}{\rho_{\rm in}}

\newcommand{\kel}{{\rm \, K}}
\newcommand{\kmps}{{\rm \: km \, s}^{-1}}
\newcommand{\nh}{n_{\rm H}}

\newcommand{\percc}{\; {\rm cm}^{-3}}

\newcommand{\alfven}{Alfv\'en}
\newcommand{\cool}{{\rm cool}}
\newcommand{\fpl}{f_{\rm PL}(T)}
\newcommand{\ftd}{f_{\rm TD}(T)}
\newcommand{\meanvs}{\left< v_s \right>}
\newcommand{\Toth}{T\'oth}

\lefthead{Kimoto \& Chernoff}
\righthead{Magnetic Fields and Overstable Shocks}

\begin{document}

\title{Effects of Magnetic Fields on Radiatively Overstable Shock Waves}

\author{Paul A. Kimoto}
\affil{Department of Physics, Cornell University, Ithaca, NY 14853}

\and 

\author{David F. Chernoff}
\affil{Department of Astronomy, Cornell University, Ithaca, NY 14853}

\begin{center}
accepted by The Astrophysical Journal \\
\copyright 1997 The American Astronomical Society
\end{center}

\begin{abstract}

  We discuss
  high-resolution simulations of one-dimensional, plane-parallel
  shock waves with mean speeds between 150 and $240 \kmps$
  propagating into gas with \alfven\ velocities up to $40 \kmps$
  and outline the conditions under which these radiative shocks
  experience an oscillatory instability in the cooling
  length, shock velocity, and position of the shock front.  
  We investigate two forms of postshock cooling:
  a truncated single power law  and a more realistic piecewise
  power law.
  The degree of nonlinearity of the instability depends strongly on the
  cooling power law and the \alfven\ Mach number: for power-law
  indices~$\alpha < 0$ typical magnetic field strengths may be insufficient
  either to stabilize the fundamental oscillatory mode or to prevent the
  oscillations from reaching nonlinear amplitudes.

\end{abstract}


\section{Introduction}

Work by Langer, Chanmugam,~\& Shaviv~(\shortcite{lcs}), Chevalier~\&
Imamura~(\shortcite{chevalier-imamura}), and Imamura, Wolff,~\&
Durisen~(\shortcite{imamura-wolff-durisen}) shows that the 
physical structure of some radiative shocks and their cooling regions is
subject to a cooling overstability.
The shock velocity and the length of the cooling column of postshock gas
may oscillate instead of evolving in a steady manner.
The linear perturbative analysis
of Chevalier~\& Imamura,
which assumes power-law 
cooling laws~$\Lambda \propto \rho^2 T^\alpha$, indicates 
that one-dimensional nonmagnetic shocks are {\em stable\/}
when $\alpha$~is sufficiently large, $\alpha \ga 0.8$.
If the cooling rises rapidly with temperature, 
perturbations to a cooling region in steady state that
increase the shock velocity~$v_s$ cause more rapid cooling and require
shorter cooling lengths, and the shock is stable against
perturbations.  On the other hand, if the cooling law does not rise
sufficiently rapidly with temperature, then the cooling length oscillates
overstably between fast shocks with long cooling lengths 
and slow shocks with short cooling lengths.

\Toth\ \& Draine (\shortcite{toth}) extend the perturbative analysis to
include a frozen one-dimensional magnetic field, 
oriented perpendicular to the gas motion and uniform in the upstream gas.  
Compression at the shock front tends to align the postshock 
magnetic field with the plane of the shock.
The field's
contribution to the total pressure 
tends to stabilize oscillations.  

In this paper we discuss sets of
one-dimensional, plane-parallel simulations of
radiative shocks, both with and without the transverse magnetic field.
Shocks in the interstellar medium propagating faster than approximately
$150 \kmps$ are candidates for these instabilities.  For example, the blast
waves of supernova remnants may pass through a velocity range in which the
shock velocities are high enough and the cooling times short enough,
compared with the remnant age, for these oscillations to occur
(\cite{kc97a}).
For cooling laws we adopt $\Lambda = \nh^2 f(T)$, where
$\nh$~is the hydrogen-nuclei density.
For the cooling function~$f(T)$ we use either (1)~single power laws or
(2)~a piecewise power-law
fit (for temperatures above $3 \times 10^4 \kel$)
to the result of Raymond, Cox, \& Smith~(\shortcite{rcs}).

We note that the latter cooling function
assumes that the emitting plasma is
in collisional equilibrium.  However, because collisional, ionization, and
recombination rates are slow compared to the cooling rate, an accurate
evaluation of the cooling necessitates a time-dependent 
treatment of the ionization evolution.  
Several specific cases of one-dimensional simulations 
incorporating explicit evolution of ionization
states are discussed by Innes, Giddings, \& Falle
(\shortcite{innes-giddings-falle}), Gaetz, Edgar, \& Chevalier
(\shortcite{gaetz-edgar-chevalier}), and Innes~(\shortcite{innes92}).  Of
these three papers, only Innes~(\shortcite{innes92}) incorporates a
transverse magnetic field.
Such treatments are computationally very demanding.
By using simpler cooling laws, we are able to investigate
a wider range of parameter space in mean shock speed and
magnetic field strength, as well as assess the sensitivity of the results
to the form of the cooling law.  
Our qualitative conclusions should
apply to more accurate treatments as well.

Our piecewise power-law cooling function differs from that of \Toth\ \& 
Draine (\shortcite{toth})
as discussed below, and
our results differ qualitatively from their similar suite of calculations.
Because our cooling function is more unstable for
mean shock velocities between $150 \kmps$ and $210 \kmps$,
we find that the stabilizing effect of magnetic fields is reduced.
In our analysis, larger magnetic fields are necessary to stabilize
large-amplitude, fundamental-mode oscillations, and as well to stabilize
all oscillatory modes.
We do not claim that our treatment
leads to results that are physically more reliable:
the differences between
the two sets of results serve to point out that quantitative
results will require a more precise treatment of the cooling.

Most recently, Walder \& Folini~(\shortcite{walder-folini}) discuss
spherically symmetric, one-dimensional, high-resolution simulations of the
radiative cooling instability.  Like ours, these
simulations use a piecewise power-law cooling function.  They
identify five qualitatively distinct oscillation types.  
We note that the two
types that we discuss in some detail below correspond to the ``strong
forms'' that they describe for the fundamental and first overtone modes.  
We do not draw further parallels with their work, however, since
their simulations do not include magnetic fields,
an essential ingredient in our discussion.

In the following section we discuss simulations carried out with our
two cooling functions in parameter spaces in which we vary the magnetic field
strength and properties of the cooling function.
We identify two distinct types of nonlinear oscillations:
one in which the fundamental mode
dominates the motion, and one in which the first-overtone mode dominates.
In the final section we summarize and draw
conclusions based on our results, taking into account the limitations of
the systems we have studied.

\section{Simulations of oscillations}

\subsection{Equations of motion and initial conditions}

In order to study the oscillatory cycles of this cooling overstability,
we consider one-dimensional, plane-parallel systems of ideal (inviscid and
with zero thermal conductivity), infinitely conducting gases, with a
magnetic field perpendicular to the direction of motion.
Our cooling law takes the form~$\Lambda = \nh^2 f(T)$.
To simplify the investigation of the oscillatory behavior, in some cases
we adopt for~$f(T)$ single power laws (with low-temperature cutoffs).
In the other cases,
as an approximation to a realistic cooling function, we use for~$f(T)$ a
piecewise power-law fit to the result of Raymond et al.~(\shortcite{rcs}).
To account for the turnoff of cooling when the
gas reaches a sufficiently low temperature to recombine, we arrange for the
cooling to vanish
below a cutoff temperature~$T_c \approx 2 \times 10^4 \kel$.
Figure~\ref{coolfcn} shows the function we adopt, as well as the function
used for application to the interstellar medium 
by \Toth\ \& Draine (\shortcite{toth}).
The points indicated in the figure denote the shock temperatures of
steady-state shocks (in the absence of magnetic fields) with velocities
$150$, $180$, $210$, and~$240 \kmps$.

Since the cooling law takes this simple functional form,
the length and time scales of solutions are simply proportional to
$1/\nh^{\rm in}$, and all densities, pressures, and energies proportional
to $\nh^{\rm in}$.
Because of this scaling, the particular value we choose for the upstream
density is arbitrary, and so usually we quote values in terms
of~$\nh^{\rm in}$ (omitting the superscript ``in'' where there is no
ambiguity).

Our numerical calculations use an Eulerian finite-difference algorithm that
can evolve flows with strong shocks and gradients.  We use operator
splitting to separate the cooling from the magnetohydrodynamics, in which
we evolve four conserved densities ($\rho$, $\rho v$, $\etotal$, and $B$)
with a flux-corrected transport scheme (\cite{zalesak}).
A multigrid method using some ideas of Berger~\&
Colella~(\shortcite{berger}) allows us to place high resolution on localized
parts of the flow as required for numerical accuracy.  Details of the
methods may be found in the Appendix of Kimoto~\&
Chernoff~(\shortcite{kc97a}).

In all of the simulations,
supersonic gas of temperature~$2500 \kel$ flows uniformly into one end of
the computational domain and is shocked.  
This gas is assumed to
be completely preionized by the photons emitted near the shock
and to contain helium in a 1:10 ratio to hydrogen.
(In our simulations, we give the
inflowing gas a hydrogen-nuclei density of $\nh^{\rm in} = 50 \percc$, but
because of the scaling properties of our equations of motion, the particular
value has no significance.)  Although the temperature of our unshocked gas
is unrealistic both for preionized gas and
for a quiescent interstellar medium, the dynamics of the shock
and the cooling column is unaffected as long as the shock is strong.  The
gas cools and accumulates near a wall at the other end, where we impose a
perfectly reflecting boundary condition.  
When the cooling region evolves in a steady state, 
the quiescent, fully cooled gas builds up at the wall, 
and the cooling region moves outward at constant velocity.  
We find considerably more complicated dynamics.

The initial conditions are constructed to correspond closely with
the steady-state cooling flow for a given shock speed and
upstream \alfven\ velocity.
The main ingredient is a numerical solution to the
equations of motion, generated by working in the shock frame and assuming
no time dependence.
First, this steady-state solution is truncated at a point
where the gas has approximately attained its final cooled,
dense state (beyond the main part of the cooling column), 
and second, the transition between upstream and shocked gas is
smoothed slightly.  The numerical evolution scheme smooths discontinuities
over several gridpoints.  By necessity the smoothing that we employ in
generating initial conditions differs slightly from the smoothing developed
under the numerical scheme.  We then transform the solution into the frame
of the cold, dense gas that builds up at the wall boundary.
In most cases the inflow velocity is only slightly less 
than the average shock speed since the gas compresses strongly.
During the highly nonlinear oscillations the conditions of the
cooling column do not resemble those of the steady state, 
and waves are driven downstream into the cooled, dense gas.
We find, however, that the measured average shock propagation speed
fluctuates only slightly about the shock speed of the steady-state initial
condition.  Table~\ref{average-vs} lists the results for twenty cases
discussed below---covering mean shock speeds from $150$~to~$240 \kmps$ and
upstream \alfven\ velocities up to~$20 \kmps$---that use the piecewise 
power-law cooling function.

\subsection{Parameter space and instability types}

The linear perturbative analysis of Chevalier~\&
Imamura~(\shortcite{chevalier-imamura}),
which assumes $f(T) \propto T^\alpha$ and no magnetic field, 
yields a discrete spectrum of modes for motions of the cooling column.
The fundamental mode has frequency
\begin{equation}    \label{fundamental-frequency}
  \nu_F \approx \frac{v_s}{21 \, L_\cool},
\end{equation}
and the overtone
frequencies are approximately odd multiples of the fundamental.  The
overtone modes are unstable over a greater range of cooling power laws than
is the fundamental.  

A transverse magnetic field can stabilize the oscillations.  The 
perturbative analysis
of \Toth~\& Draine~(\shortcite{toth}), which incorporates such a magnetic
field, uses
a cooling function
\begin{equation} \label{td-cooling}
  \ftd \propto \cases{T^\alpha&if $T > T_t$,\cr
                      T^{1/2}&if $T < T_t$,\cr
                     }
\end{equation}
for some turnover temperature~$T_t$.
For shock temperatures much greater than $T_t$, the minimum stabilizing
magnetic field depends mainly on~$\alpha$: the field
is greater for smaller values
of~$\alpha$ (i.e., the more unstable cases), and it is greater for overtone
modes than for the fundamental.  In addition, the stabilizing field for all
overtone modes is approximately the same.
In nonlinear calculations typically only
the first overtone mode is observed.
The results can be summarized as follows:
the minimum stabilizing field for mode~$i$ may be parameterized
by~$\machrecipcrit$ (where the \alfven\ Mach number is~$M_A = v_s/v_A$, the
\alfven\ velocity is~$v_A = B/(4 \pi \rhoIN)^{1/2}$, and the upstream
density is $\rhoIN$).
For small magnetic fields, $\machrecip < \machrecipfund < \machrecipover$,
both fundamental and overtone modes are unstable;
for intermediate fields, $\machrecipfund < \machrecip < \machrecipover$,
only the overtones are unstable;
and for large fields, $\machrecipfund < \machrecipover < \machrecip$,
all modes are stable.
The short dotted lines in Figure~\ref{m-alpha} show approximately
the~$\machrecipfund$ (lower) and~$\machrecipover$ (upper) curves 
in the $(\alpha, \machrecip)$ parameter space as calculated by \Toth\ \&
Draine using the cooling function~(eq.~[\ref{td-cooling}]) with turnover
temperature~$T_t = 10^{-3} \mu v_s^2/k_B$, where~$\mu$ is the mean particle
mass and $v_s$~is the mean shock speed.

\subsection{Oscillations without magnetic fields}

As background, we begin by describing properties of oscillations 
in the absence of magnetic fields.  
For simulations adopting the piecewise power-law cooling function,
Table~\ref{oscillations} 
lists typical amplitudes (minimum to maximum shock velocities) 
and periods for a range of shock speeds---from 130 to $240 \kmps$---that 
should cover a range of supernova remnants
with oscillating shocks (\cite{kc97a}).
The corresponding range of shock temperatures covers
a break in the power-law fit (locally proportional to $T^\alpha$)
that enters our cooling function: 
Above $5 \times 10^5 \kel$, which corresponds
to a $190 \kmps$ shock (neglecting magnetic fields), $\alpha \approx -0.1$;
from $2.5 \times 10^5$ to $5 \times 10^5 \kel$, we 
adopt $\alpha \approx -2.2$ (for which the oscillations are more unstable).

From these simulations we make several observations
that are generally consistent with the results of previous work done with
a variety of steep cooling laws (e.g., \cite{imamura-wolff-durisen};
\cite{innes-giddings-falle}; \cite{gaetz-edgar-chevalier};
\cite{walder-folini}).
The instability is so strong that within a few periods 
the oscillations develop and their amplitudes quickly saturate.  
When the oscillations become nonlinear, the fundamental mode dominates.
Our amplitudes and periods typically vary by up to~$10\%$,
but the values indicated in Table~\ref{oscillations}
should be representative.
The period of the fundamental mode is close to
$\tau_F = 21 \, L_\cool/\meanvs$, approximately the value
(cf.\ eq.~[\ref{fundamental-frequency}])
predicted by linear analysis (\cite{chevalier-imamura}).
We show this approximate value for comparison in Table~\ref{oscillations}, 
where (in the spirit of small-amplitude analysis)
we take $L_\cool$ from the steady-state solutions.
(In the perturbative treatments the gas cools to~$T= 0$ 
[for analytic convenience];
our cooling function turns off in a tapered manner at a finite temperature 
[in part, for numerical convenience].  Because the gas cools rapidly as it
reaches high densities, it is still possible to identify a ``cooling
length'' without much ambiguity.)
In our simulations the oscillations become highly nonlinear, and the
oscillatory periods are somewhat longer than the linear-analysis values.

\subsection{Oscillations affected by magnetic fields}

Next we consider simulations with a transverse magnetic field.
For simplicity we begin by considering
cooling functions with single power laws in order 
to investigate the effects of power-law slope and magnetic field on
the selection of the dominant mode.
For a cooling function
(cf.~eq.~[\ref{td-cooling}]) we use
\begin{equation}  \label{pl-cooling}
  \fpl \propto \cases {T^\alpha&if  $T > T_t$,\cr
                       T^{1/2} 
                        \left\{
                          \frac{1}{2} +
                            \frac{1}{2}
                            \tanh [(T-\frac{9}{14}T_t)/(\frac{1}{49} T_t)]
                        \right\}&if $T < T_t$.\cr
                      }
\end{equation}
The quantity in braces arranges for the cooling to effectively vanish at
small temperatures ($T < 9T_t/14$).
For~$\alpha \ge -0.5$
we (like \Toth\ \& Draine~[\shortcite{toth}]) set the turnover temperature 
so that~$k_B T_t/(\mu v_s^2) = 10^{-3}$.
For~$\alpha \le -0.8$, however,
we use~$k_B T_t/(\mu v_s^2) = 10^{-2}$ for numerical convenience.

\begin{figure}
\myepsf{6.0in}{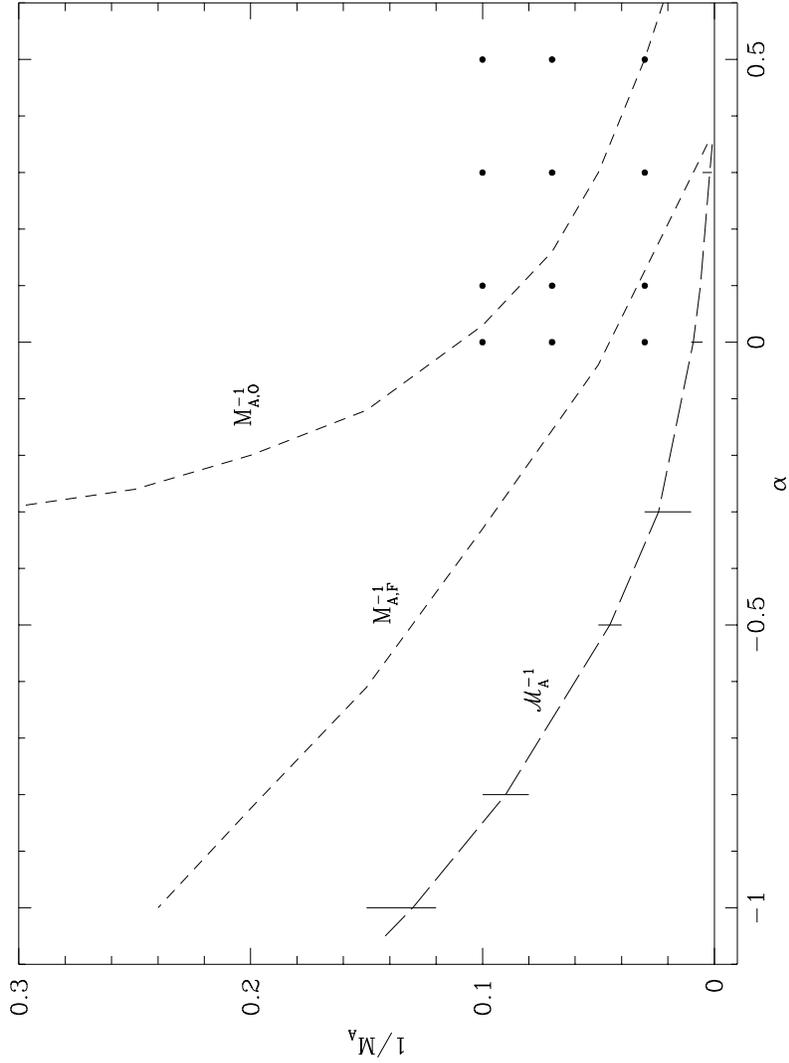}{0}
\caption{From our simulations with power-law cooling,
we find that the fundamental mode dominates the
instability roughly in the region below the long dashed lines in 
($\alpha, \machrecip$) parameter space, where~$\alpha$ is the power-law
index.  (The short
line segments indicate simulations between which the fundamental mode becomes
stable.)
The short dotted lines indicate curves of marginal stability for the
fundamental~($\machrecipfund$, lower) and overtone~($\machrecipover$, upper)
modes as calculated by \Toth\ \& Draine (\protect\shortcite{toth}).
The dots indicate the twelve simulations discussed by \Toth\ \& Draine.
\label{m-alpha}}
\end{figure}

In the previous subsection we noted that the fundamental mode
dominates the oscillations in the absence of magnetic fields 
($\machrecip = 0$).
At several values of the power-law
index~$\alpha$ (namely, $-1$, $-0.8$, $-0.5$, $-0.3$, 0, and 0.3),
we bracket the
reciprocal \alfven\ Mach number~$\machreciptrans$ at which dominance switches
from the fundamental to an overtone mode.  Simulations yield the limits in
$\machreciptrans$ shown as line segments in Figure~\ref{m-alpha}, and the
curve in long-dashed lines (drawn only for illustrative purposes) shows
approximately the behavior of~$\machreciptrans$.  Naturally 
this curve must lie below
the lower short-dashed line, which denotes~$\machrecipfund$, the value at
which the fundamental mode is marginally stable in the linear analysis of
\Toth\ \& Draine.  We note that the magnitude of $\machreciptrans$ curve is
clearly distinct from that of $\machrecipfund$.

\begin{figure}
\myepsf{6.0in}{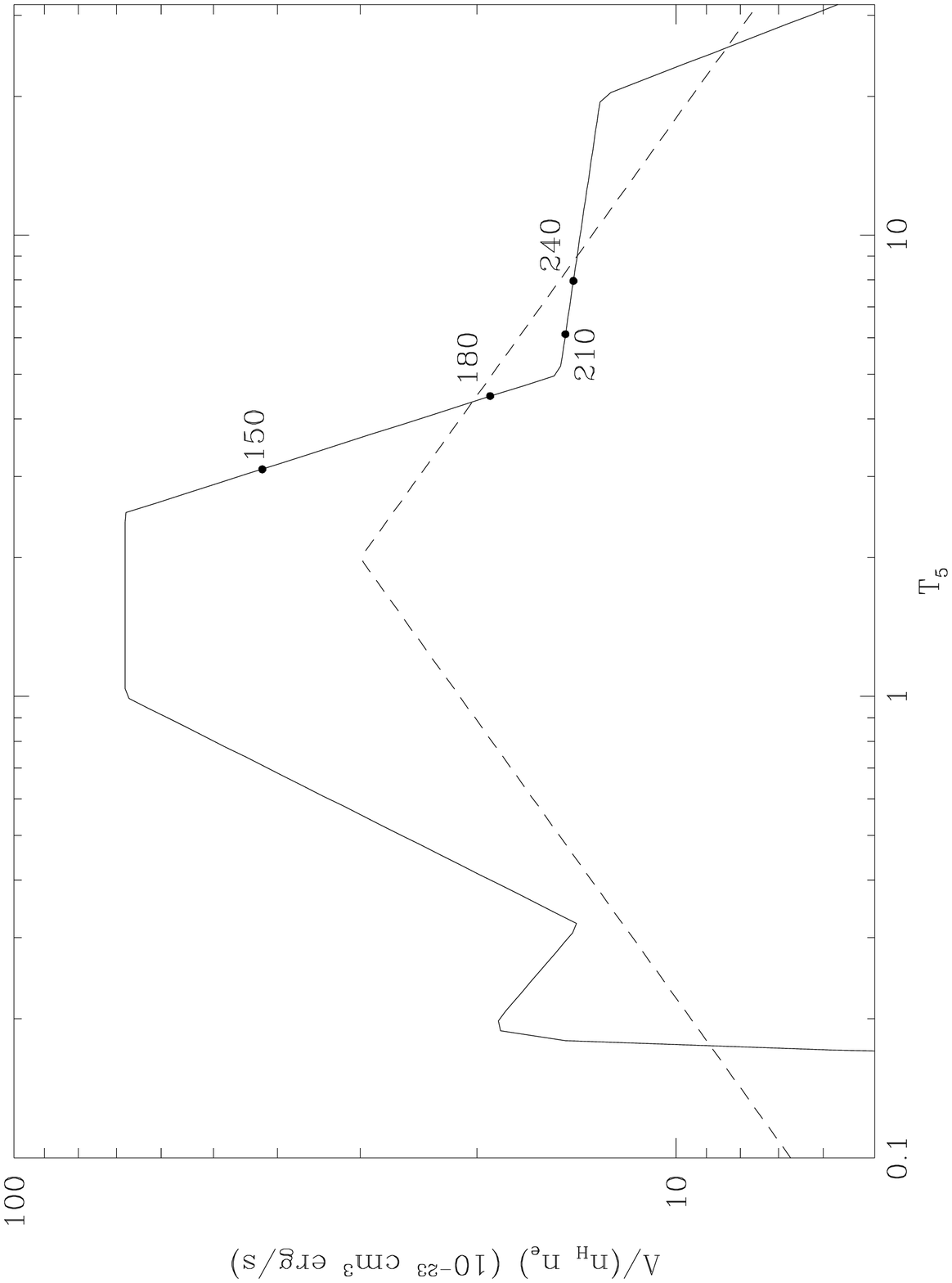}{0}
\caption{The simulations use the cooling function indicated by the solid
line, a piecewise power-law fit 
to the result of Raymond et al.~(\protect\shortcite{rcs}), 
and arranged to turn off a low temperatures.  The dashed line shows the
simple function adopted by \Toth\ \& Draine (\protect\shortcite{toth}) for
application to the interstellar medium.  The four dots indicate the
postshock gas temperatures for shocks moving at $150$, $180$, $210$, and
$240 \kmps$.
\label{coolfcn}}
\end{figure}

\begin{figure}
\myepsf{6.0in}{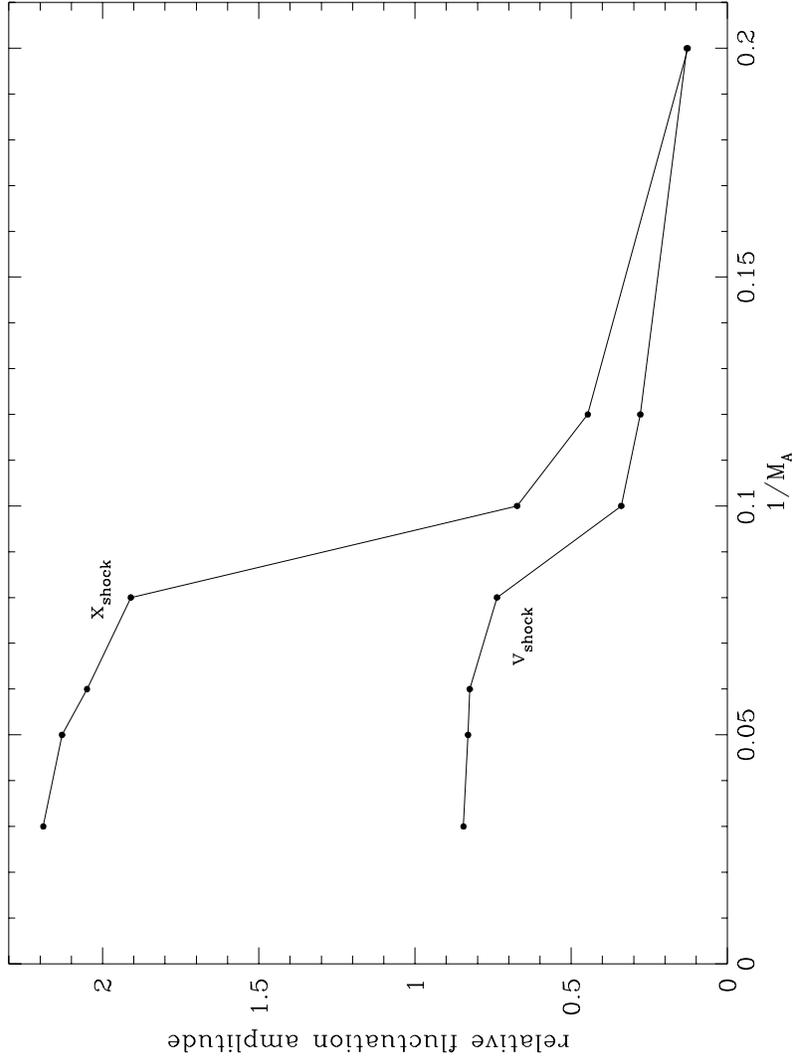}{0}
\caption{Typical sizes of shock fluctuations (relative to steady-state
  values) as a function of reciprocal \alfven\ Mach number, for
  power-law cooling~(eq.~[\protect\ref{pl-cooling}]) with~$\alpha = -0.8$.
  The fundamental mode dominates for~$\machrecip \le 0.08$;
  and the overtone, for~$\machrecip \ge 0.1$.
  \label{avg-amplitude}}
\end{figure}

\Toth\ \& Draine discuss the results of hydrodynamical simulations with the
values of $\alpha \ge 0$ and $\machrecip \ge 0.03$ denoted by dots in
Figure~\ref{m-alpha}.  They observe that
the first-overtone mode dominates whenever the flow is
unstable and that the oscillations have amplitudes~$\la \! 10\%$.
We have performed several simulations in the vicinity of the dots using
our version of the power-law cooling function~(eq.~[\ref{pl-cooling}]), 
and these
confirm their conclusions.  However, Figure~\ref{m-alpha} also shows that
{\em the fundamental mode dominates the oscillations over an
astrophysically significant range of magnetic fields when~$\alpha < 0$.}
Even the simplest
forms of cooling functions (e.g., as shown in Fig.~\ref{coolfcn}) require
temperature ranges in which locally~$\alpha < 0$.
In addition, for $\alpha < 0$ the oscillations are decidedly
nonlinear: when the fundamental mode dominates, the velocity fluctuations
(minimum to maximum) are at least 50\% of the mean shock velocity.
In the
handful of cases we have simulated in which the overtone mode dominates
(i.e., with~$\machrecip$ slightly greater than~$\machreciptrans$),
the velocity fluctuations are typically~$\sim \! 25\%$.
Figure~\ref{avg-amplitude} shows the magnitude of
fluctuations in cooling-column length and shock velocity 
(relative to the steady-state 
values) as a function of~$\machrecip$ when~$\alpha = -0.8$.
When the fundamental mode dominates, as occurs for~$\alpha \la 0.08$, the
fluctuations are very strong.  The amplitudes drop abruptly when the overtone
mode becomes the dominant mode (at~$\machrecip \approx 0.1$)  All our
numerical evidence confirms that the flows have strong,
nonlinear fluctuations when the physical parameters lie
in the region below~$\machreciptrans$ in Figure~\ref{m-alpha}.  
Above~$\machreciptrans$ the
fluctuation amplitudes, while smaller than below, may still be appreciable.
The parameter space covered by the simulations of \Toth\ \& Draine does
not cover
the full range of plausible cooling-law slopes,
and consequently several conclusions for our more realistic
cooling function contradict their general results.

We end our discussion of single power-law cooling functions
with comments about the robustness of the values of~$\machreciptrans$ 
with respect to changes in simulation parameters.
As noted above, in the cases with~$\alpha \le -0.8$,
for numerical convenience we use a value of the turnover temperature~$T_t$
ten times larger than the higher-$\alpha$ cases.
We have verified in the case of~$\alpha = -0.5$ that the selection of the
dominant mode discussed below is insensitive to this increase in~$T_t$.
For smaller values of~$\alpha$, we are interested in the behavior at
increasingly large magnetic fields.  The effect on the dynamics of the
cooling column should be
negligible since $T_t$~sets the thermal pressure of the cooled gas, 
but near the mode transition this thermal pressure is dominated by the
magnetic pressure.  For larger values~$\alpha \ga 0$,
cases with small or zero magnetic fields are of interest, 
and the choice of final gas temperature does qualitatively affect the
oscillations. 

We believe that our choice of boundary conditions has little effect on the
value of~$\machreciptrans$.
However, simulations of
one-dimensional cooling shocks without magnetic fields by Strickland \&
Blondin~(\shortcite{strickland}) suggest that boundary conditions may affect
the stability or the dominant unstable mode.  They offer as a possible
explanation that a reflecting
wall may return waves to the cooling region and interfere with the growth
of unstable modes.  We use
the case of~$\alpha = -0.5$ to check the dependence of~$\machreciptrans$
on boundary conditions.
In order to remove the effect of reflected waves, we separate the cooling
region from the wall with a column of quiescent gas large enough so that
waves generated in the region of the shock are not reflected back to the
cooling column during the simulation.  We find no change in the qualitative
behavior of cases on either side of our inferred $\machreciptrans$ boundary.

\begin{figure}
\myepsf{6.0in}{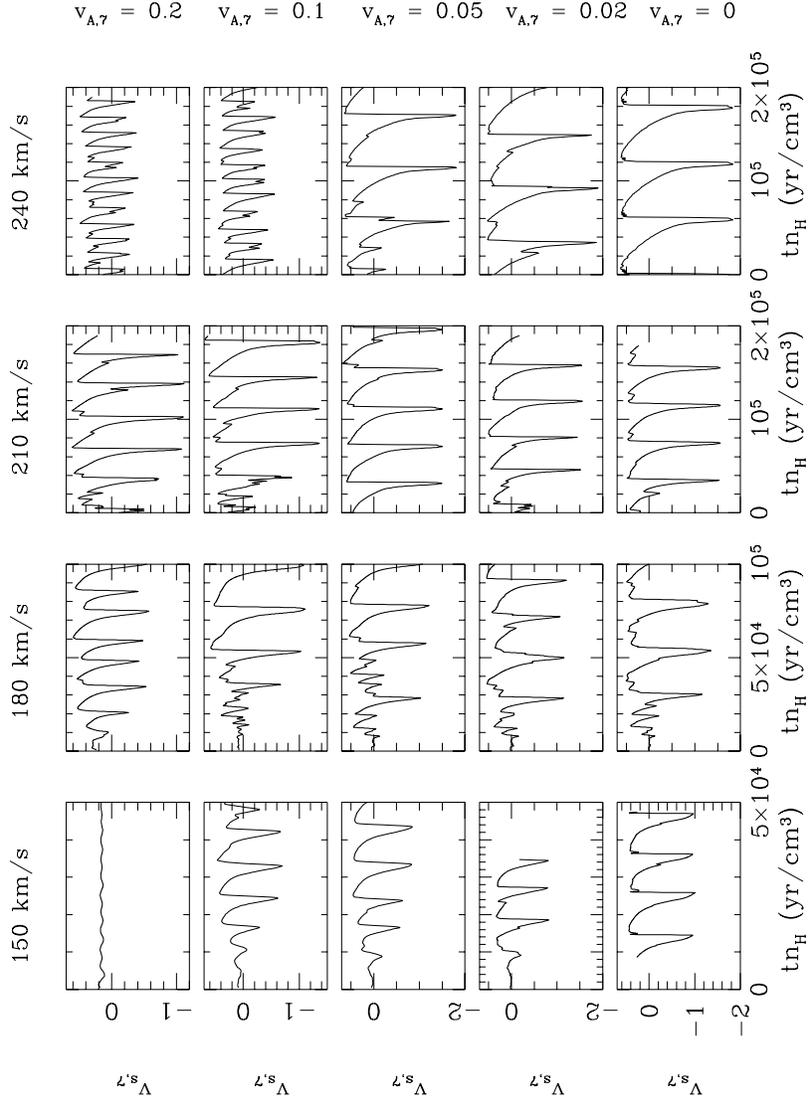}{0}
\caption{Shock velocity (in units of $100 \kmps$) as a function of time 
for mean shock speeds
150, 180, 210, and $240 \kmps$ (from left to right) and upstream \alfven\
velocities 0, 2, 5, 10, and $20 \kmps$ (from bottom to top).
\label{vs-mosaic}}
\end{figure}

Next we return to the more complex piecewise power-law cooling function shown
in Figure~\ref{coolfcn}.  In Figure~\ref{vs-mosaic} we show the
shock-velocity fluctuations for the cases
mean shock speeds 150, 180, 210, and $240 \kmps$, each with \alfven\
velocities 0, 2, 5, 10, and $20 \kmps$.  The graphs show the shock velocity
(in the frame of reference of the reflecting wall) as a function of time.
Over this range of shock speeds, all $B=0$ flows are unstable.  Since the
local slope of the cooling function varies, the lower shock speeds should be
more unstable than the higher ones.  We discuss the effects on stability,
dominant mode, and oscillation amplitude as the magnetic field is
increased.

Inspection of Figure~\ref{vs-mosaic} shows that (1)~the fundamental mode
dominates over most of the parameter space, (2)~the amplitude of
oscillations is nonlinear~($\sim \! 100\%$), and
(3)~the magnetic fields qualitatively alters the instability only at the
lowest and highest shock speeds.
In a few cases starting from near steady-state conditions,
the growth rate of an unstable fundamental mode is relatively
slow, and the qualitative long-term behavior sets in only after
10--20 cyclic times.  Before this occurs, the shock velocity oscillates
with ever-increasing amplitude, and the variations in shock velocity are
dominated by the overtone modes.
We now discuss these observations in light of the conclusions
drawn from the single power-law cooling simulations described above.

In two high-speed cases 
($v_s = 240 \kmps$, $v_A = 10 \, \mbox{and} \, 20 \kmps$),
the magnetic field is sufficiently strong that the fundamental, but
not the overtone, mode is stabilized.
Under these circumstances the oscillations are noticeably
smaller in amplitude than in the corresponding cases with lower magnetic
fields.  
As expected for the first-overtone mode, the oscillation frequency is
roughly three times that of the lower-magnetic-field 
$240 \kmps$ cases, whose fundamental modes are unstable.
(At high \alfven\ Mach numbers
the contribution of the magnetic field to the total pressure is
small in gas that has not yet become cool and dense, and
the cooling length and mode frequencies are not much altered from the
$v_A = 0$ case.) 
The simulations show that~$0.02 < \machreciptrans < 0.04$.

We may compare this with the results obtained using single power-law cooling.
The relevant parameter for comparison is the power-law index~$\alpha$.
At the shock temperature corresponding to $240 \kmps$, our
cooling function locally has~$\alpha \approx -0.1$.  In that case,
our power-law results (cf.~Fig.~\ref{m-alpha})
suggest that~$\machreciptrans \approx 0.02$, roughly consistent
with the $240 \kmps$ simulations.  The value
of~$\machreciptrans$ is likely increased because the cooling function
is
steeper ($\alpha \approx -2.2$) in a wide range of temperatures below
the shock temperature.  

In one case ($v_s = 150 \kmps$, $v_A = 20 \kmps$) in
Figure~\ref{vs-mosaic}, the instability is absent; 
the oscillations shown are extremely small in amplitude and decaying.
The simulations show that $0.07 < \{\machrecipfund, \machrecipover\} < 0.13$.
At a shock speed of~$150 \kmps$, the shock temperature---if $B = 0$---implies
a local slope~$\alpha \approx -2.2$.
Although such a steep power law is not considered by the linear analysis of
\Toth\ \& Draine,
extrapolation from their results indicates that~$\machrecip > 0.4$ would be
required to stabilize the fundamental mode.  
In this case the observed stabilization is
of a different nature: the upstream magnetic field is sufficiently large
to reduce the shock temperature by more than 10\% 
as compared with the $B=0$~case.  At this reduced temperature 
the cooling function is shallow ($\alpha = 0$) and less unstable.
The combination of the curvature of the cooling function
and the lowered shock temperature stabilizes
fundamental and overtone modes in
part of parameter space.

\begin{figure}
\myepsf{6.0in}{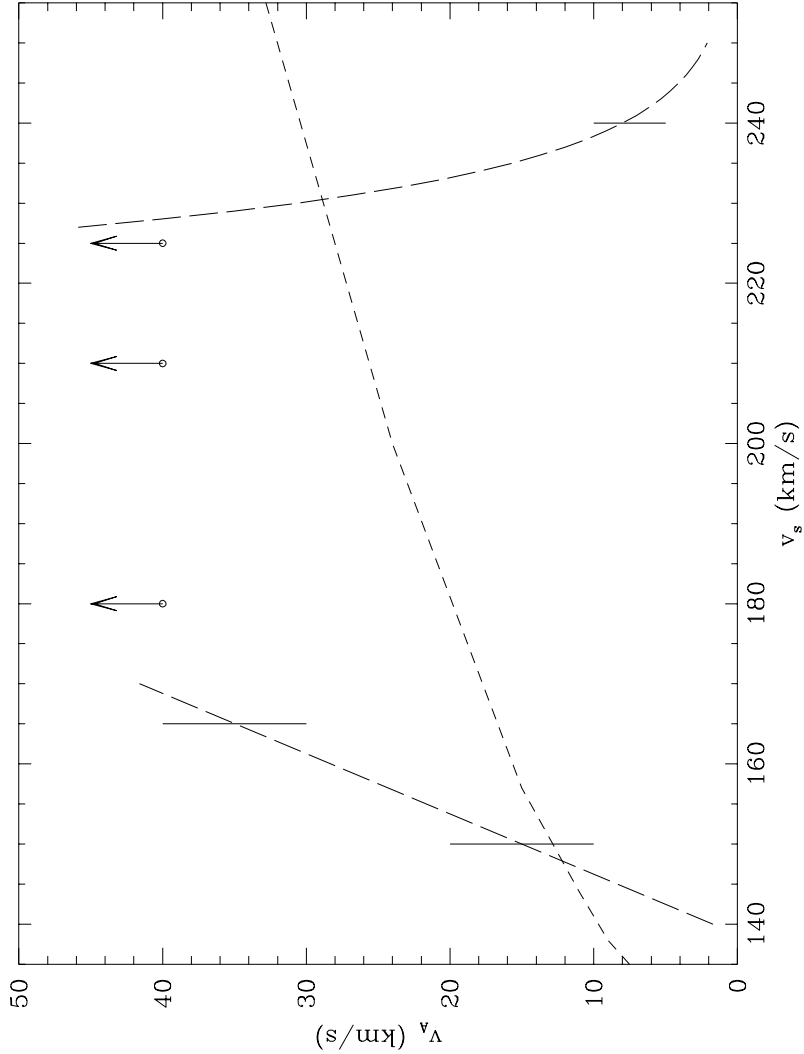}{0}
\caption{From our simulations we find the fundamental mode is unstable
roughly in the region below the long dashed lines.  (The short line
segments indicate simulations between which the fundamental mode becomes
stable; the transition happens at magnetic fields above the points marked
with arrows.)
For comparison, the
region below the short dashed lines is found to be unstable to the
fundamental mode by \Toth\ \& Draine (\protect\shortcite{toth}).  The
difference is attributable to the different cooling function
(cf.~fig.~\protect\ref{coolfcn}).
\label{paramsp}}
\end{figure}

Finally, we come to the issue of stability over the large central range of 
shock speeds.  We see no evidence of stabilization 
at $v_s = 180$~and~$210 \kmps$
for~$v_A \le 20 \kmps$ in Figure~\ref{vs-mosaic}.  We have performed
additional simulations with larger magnetic fields 
($v_A = 30$~and~$40 \kmps$), and the results for the fundamental
mode are summarized in Figure~\ref{paramsp}.  
The vertical line segments
indicate constraints on the boundary of the fundamental-mode dominance.
Arrows indicate that fundamental-mode stability requires {\em larger\/}
magnetic fields than those we simulated.  
In the region below the lines marked in long dashes
(drawn only for illustrative purposes) the fundamental mode dominates.

This picture is qualitatively different from that drawn by \Toth\ \&
Draine.  For a very rough approximation to the interstellar-medium cooling
function, they consider the simple broken power law shown in
Figure~\ref{coolfcn}.  Their linear analysis then yields for the
fundamental-mode stability limit the line drawn in short dashes in
Figure~\ref{paramsp}.  The results for our cooling function indicate that
{\em much larger\/} magnetic fields are required to change the stability of
the oscillations.
The major difference between the cooling functions is 
that ours is much steeper (and hence more unstable) for shocks with 
$v_s \le 200 \kmps$, but flatter 
(hence more stable) for higher-velocity shocks.  Our picture agrees with
the results from single power-law cooling above:
the fundamental mode dominates the oscillations, and the oscillations have
significant amplitudes in the presence of typical magnetic fields 
when $\alpha < 0$.

\section{Conclusions} \label{conclusions}

In our simulations the behavior of the overstability is determined by the
typical shock temperature (hence the mean shock speed) and the upstream
magnetic field.  We divide the unstable cases into two categories: either the
fundamental mode is unstable, or it is absent and the instability is
dominated by the first overtone mode.  
The periods of these observed modes are roughly consistent with the
prediction of linear perturbative analyses (\cite{chevalier-imamura}; 
\cite{toth}).
Simulations with cooling functions~(eq.~[\ref{pl-cooling}]) 
that are truncated one-component power laws show a strong
dependence on power-law index~$\alpha$ and the \alfven\ Mach
number~$\machrecip$.  The earlier simulations presented in \Toth\ \&
Draine~(\shortcite{toth}) all have~$\alpha \ge 0$.
For typical magnetic field strengths they found
oscillations of limited~($\la 10\%$) amplitudes dominated by overtone modes.
For similar magnetic fields but~$\alpha < 0$,
we observe that the fundamental mode dominates 
and that the oscillations have large amplitudes. 
We note a significant drop in oscillation amplitude
accompanying the switch from fundamental to overtone mode as the magnetic
field is raised, 
as we show for the typical case of~$\alpha = -0.8$.  
Even when the overtone mode dominates, the
oscillations have substantial, nonlinear amplitudes.  
The linear analysis of \Toth\ \& Draine
yields the values of the reciprocal Mach numbers~$\machrecipfund$
and~$\machrecipover$ at which the fundamental and overtone modes become
linearly stable for the full range of~$\alpha$.  
Our nonlinear simulations do not disagree with the linear analysis:
the reciprocal Mach
number~$\machreciptrans$ at which the dominant oscillations 
change from fundamental
to overtone mode differs significantly from~$\machrecipfund$, but always
satisfies~$\machreciptrans < \machrecipfund$.

These single power-law cooling results explain our simulations that use
a more realistic, piecewise power-law cooling function based on the
results of Raymond et al.~(\shortcite{rcs}).
These simulations encompass mean shock
speeds between $150$ and $240 \kmps$ and \alfven\ velocities up to $20
\kmps$.  Over most of the parameter space we observe large-amplitude
fundamental-mode oscillations.  The instability is so strong that quite
substantial magnetic fields ($v_A \ga 10 \kmps$)
are required to change the qualitative behavior,
and then only at the lowest and highest shock speeds.
Again, we contrast our qualitative picture with the discussion of
\Toth\ \& Draine, who 
consider a very rough model for the interstellar-medium cooling function with
a single power law ($\alpha = -1/2$) above~$T = 10^{5.3} \kel$.
With a linear
analysis they find that magnetic fields of realistic amplitudes may stabilize
radiative shocks with velocities up to
approximately 
$175 \kmps$ in the warm ionized medium and the warm neutral medium.
For some cases, we find that fields {\em larger\/} than any we
simulated---that is, $v_A > 40 \kmps$---would be
required to affect the stability behavior.  The
discrepancy between the two treatments
points out that an accurate determination of the magnetic field
required to stabilize shocks requires a more careful treatment of the
cooling than the one we have attempted (see, for example, \cite{innes92}).
However, the minimum magnetic field to stabilize the shock is
quite dependent on the local power-law index~$\alpha$ 
of the cooling function, and in any realistic
treatment we expect~$\alpha$ to vary over the velocity range of interest
($130 \kmps \la \left< v_s \right> \la 250 \kmps$).  We then anticipate that
the stabilizing magnetic field will vary from case to case
but could be quite large, as found in our simulations.

It must be noted that some aspects of gas behavior 
are not well modeled in our simulations.
Perhaps the most prominent missing element is 
an accurate treatment of cooling, incorporating
ionization and recombination in the
cooling region, as well as in the upstream preshock gas and the cooled
postshock gas.
In particular, our
cooling function only crudely mimics the turnoff of the cooling law upon
recombination.  There may as well be geometrical effects that require
extending this work into two or three spatial dimensions.  Further work
will need to address one or more of these deficiencies.

\acknowledgements

We thank Edwin Salpeter and the anonymous referee for useful comments on
drafts of this paper.

This research has been carried out at Cornell University with the generous
support of the NSF~(AST 91-19475) and NASA~(NAGW-2224)
under the LTSA~program.  

Some computations reported herein were carried out using the resources of
the Cornell Theory Center, which receives major funding from the~NSF and
New York State, with additional support from~ARPA, the National Center for
Research Resources at the~NIH, IBM Corporation,
and other members of the Center's Corporate Partnership Program.

\pagebreak

\pagebreak  

\begin{table}

\caption{Average shock speeds in simulations
\label{average-vs}}

\begin{tabular}{rrrr}
\tableline
\tableline
Intended average & Upstream \alfven\ velocity & Inflow velocity &
Actual average (relative) \\
($\kmps$) & ($\kmps$) & ($\kmps$) & ($\kmps$) \\ \tableline
150 & 0   & 149 & $150 \pm 1$ \\
150 & 2   & 148 & $149 \pm 1$ \\
150 & 5   & 146 & $151 \pm 3$ \\
150 & 10  & 142 & $150 \pm 1$ \\
150 & 20  & 134 &  150 \\ 
180 & 0   & 179 & $181 \pm 1$ \\
180 & 2   & 178 & $182 \pm 2 $ \\
180 & 5   & 176 & $181 \pm 1$ \\
180 & 10  & 172 & $180 \pm 2 $ \\
180 & 20  & 157 & $177 \pm 2$ \\
210 & 0   & 209 & $210 \pm 1$ \\ 
210 & 2   & 208 & $210 \pm 1$  \\
210 & 5   & 206 & $210 \pm 1 $ \\
210 & 10  & 202 & $211 \pm 3$ \\
210 & 20  & 195 & $210 \pm 2 $ \\
240 & 0   & 239 & $240 \pm 1$ \\
240 & 2   & 238 & $238 \pm 2$ \\
240 & 5   & 236 & $238 \pm 3 $ \\
240 & 10  & 232 & $240 \pm 1$ \\
240 & 20  & 225 & $240 \pm 2$ \\
\tableline
\end{tabular}

\end{table}

\begin{table}

\caption{Typical parameters of shock-velocity oscillations 
without magnetic fields\label{oscillations}}

\begin{tabular}{rrrr}
\tableline
\tableline
Mean shock velocity & 
 Velocity fluctuation & 
 Fluctuation period\tablenotemark{a} &
 Linear period $\tau_F$ \\
($\kmps$) & ($\kmps$) & (yr) & (yr) \\ \tableline
130 &  80 &  4 400 & 4 000 \\
140 &  95 &  7 000 & 4 500 \\
150 & 140 & 11 000 & 5 600 \\
160 & 140 & 13 000 & 7 900 \\
170 & 160 & 17 000 & 11 000 \\
180 & 160 & 20 000 & 16 000\\
190 & 160 & 30 000 & 22 000 \\
200 & 160 & 34 000 & 29 000 \\
210 & 200 & 40 000 & 35 000 \\
225 & 210 & 47 000 & 43 000 \\
240 & 240 & 63 000 & 53 000 \\
\tableline
\end{tabular}

\tablenotetext{a}{$n_H = 1 \percc$; all times vary as $n_H^{-1}$}

\end{table}


\newpage

\begin{thebibliography}{}

\bibitem[\protect\citename{Berger~\& Colella~}1989]{berger}
  Berger, M. J., \& Colella, P. 1989, J. Comput. Phys., 82, 64

\bibitem[\protect\citename{Chevalier~\& Imamura~}1982]{chevalier-imamura}
 Chevalier, R. A., \& Imamura, J. N. 1982, \apj , 261, 543

\bibitem[\protect\citename{Gaetz, Edgar~\& Chevalier~}1988]%
{gaetz-edgar-chevalier}
  Gaetz, T. J., Edgar, R. J., \& Chevalier, R. A. 1988, \apj , 329, 927

\bibitem[\protect\citename{Imamura, Wolff,~\& Durisen~}1984]%
{imamura-wolff-durisen}
 Imamura, J. N., Wolff, M. T., \& Durisen, R. H. 1984,
              \apj , 276, 667

\bibitem[\protect\citename{Innes~}1992]{innes92}
Innes, D. E. 1992, \aap , 256, 660

\bibitem[\protect\citename{Innes, Giddings,~\& Falle~}1987]%
{innes-giddings-falle}
 Innes, D. E., Giddings, J. R., \& Falle, S. A. E. G. 1987,
              MNRAS, 226, 67

\bibitem[\protect\citename{Kimoto~\& Chernoff~}1997]{kc97a}
  Kimoto, P. A., \& Chernoff, D. F. 1997, \apj , 485, 274

\bibitem[\protect\citename{Langer, Chanmugam,~\& Shaviv~}1981]{lcs}
 Langer, S. H., Chanmugam, G., \& Shaviv, G. 1981, \apjlett , 245, L23

\bibitem[\protect\citename{Raymond, Cox~\& Smith~}1976]{rcs}
 Raymond, J. C., Cox, D. P., \& Smith, B. W. 1976,
              \apj , 204, 290

\bibitem[\protect\citename{Strickland~\& Blondin~}1995]{strickland}
  Strickland, R., \& Blondin, J. M. 1995, \apj , 449, 727

\bibitem[\protect\citename{\Toth{}~\& Draine~}1993]{toth}
 \Toth , G., \& Draine, B. T. 1993, \apj , 413, 176

\bibitem[\protect\citename{Walder~\& Folini~}1996]{walder-folini}
  Walder, R., \& Folini, D. 1996, \aap , 315, 265

\bibitem[\protect\citename{Zalesak~}1979]{zalesak}
   Zalesak, S. T. 1979, J. Comput. Phys., 31, 335

\end{thebibliography}
\end{document}